# Code Sharing in Healthcare Research: A Practical Guide and Recommendations for Good Practice

Lukas Hughes-Noehrer* (1,2), Matthew J Parkes* (3,4,5), Andrew Stewart (1), Anthony J Wilson (2,6), Gary S Collins (7,8), Richard D. Riley (7,8), Maya Mathur (9), Matthew P. Fox (10), Nazrul Islam (11), Paul N Zivich (12), and Timothy Feeney✉ (12).

1) Department of Computer Science, School of Engineering, The University of Manchester, Manchester, UK
2) Department of Anaesthesia, Critical Care and Perioperative Medicine, Manchester University NHS Foundation Trust, Manchester, UK
3) Centre for Biostatistics, Division of Population Health, Health Services Research & Primary Care, School of Health Sciences, The University of Manchester, Manchester, UK
4) NIHR Manchester Biomedical Research Centre (BRC), Manchester University NHS Foundation Trust, Manchester Academic Health Science Centre (MAHSC), UK
5) NIHR Manchester HealthTech Research Centre (HRC) in Emergency & Acute Care, Manchester University NHS Foundation Trust, Manchester Academic Health Science Centre (MAHSC), UK
6) Christabel Pankhurst Institute for Health Technology Research and Innovation, The University of Manchester, UK
7) Department of Applied Health Sciences, School of Health Sciences, College of Medicine and Health, University of Birmingham, Birmingham, UK
8) NIHR Birmingham Biomedical Research Centre, Birmingham, UK
9) Quantitative Sciences Unit and Department of Pediatrics, Stanford University, Stanford, CA, USA
10) Departments of Epidemiology and of Global Health, Boston University, MA
11) School of Primary Care, Population Sciences and Medical Education, Faculty of Medicine, University of Southampton, Southampton, UK
12) Department of Epidemiology, Gillings School of Public Health, University of North Carolina at Chapel Hill, Chapel Hill, NC, USA

Correspondence to: T Feeney, feeney@unc.edu ORCID: 0000-0001-8139-5830

**Funding:**

Dr. Feeney is supported by a National Institute of Allergy and Infectious Diseases grant R01AI157758. Dr. Zivich was supported by a developmental award from National Institute of Allergy and Infectious Diseases (K01AI177102). Dr Hughes-Noehrer is supported by an Engineering and Physical Sciences research grant EP/X02945X/1. Dr Parkes is supported by the National Institute for Health and Care Research (NIHR) Manchester Biomedical Research Centre (BRC) (NIHR203308), and the NIHR HealthTech Research Centre (HRC) for Emergency and Acute Medicine (NIHR205301). Dr. Mathur was supported by National Institutes of Health grants R01LM013866, UL1TR003142, P30CA124435, and P30DK116074. Professor Collins and Riley are National Institute for Health and Care Research (NIHR) Senior Investigators. Prof Riley is supported by the National Institute for Health and Care Research (NIHR) Birmingham Biomedical Research Centre at the University Hospitals Birmingham NHS Foundation Trust and the University of Birmingham.

Funders didn't influence the results/outcomes of the study despite author affiliations with the funder. The views expressed are those of the authors and do not necessarily represent those of the National Institutes of Health, NIHR, the Department of Health and Social Care, or UKRI.

**Competing interests:**

Timothy J Feeney has received consulting fees from the BMJ in his functions as Research Editor for the BMJ and Clinical Editor for BMJ Medicine.

**Contributors:**

Conceptualisation: LHN; MJP; TF

Writing - Original Draft: LHN; MJP; TF

Writing - Review & Editing: LHN; MJP; AS; AW; GSC; RDR; MM; MPF; NI; PNZ; TF

*LHN & MJP contributed equally to this paper.

**Standfirst:**

*As computational analysis becomes increasingly more complex in health research, transparent sharing of analytical code is vital for reproducibility and trust. This practical guide, aligned to open science practices, outlines actionable recommendations for code sharing in healthcare research. Emphasising the FAIR (Findable, Accessible, Interoperable, Reusable) principles, the authors address common barriers and provide clear guidance to help make code more robust, reusable, and scrutinised as part of the scientific record. This supports better science and more reliable evidence for computationally-driven practice and helps to adhere to new standards and guidelines of codesharing mandated by publishers and funding bodies.*

**Introduction and motivation:**

Analysis of health research is increasingly complex. In addition to decision-making about study conduct, ethics, and monitoring, there are numerous decisions made about how the analysis of a study is conducted, each of which can have material effects on the results[1–3]. However, despite numerous reporting guidelines for various study designs, many of these decisions, and the rationale behind them, are either reported at a cursory level in the methods section of the study manuscript, or omitted entirely[4–6]. Word limits, technical complexity, and the desire for manuscripts to have a 'narrative' can encourage writers to reduce the methodological content of a study report for the sake of brevity. Statistical analysis plans (SAPs) typically contain more detail on the planned analysis process and the underpinning decision-making, but these are rarely published or even written for some types of clinical research (e.g., observational studies), and there is no requirement to produce this documentation. While detailed, a SAP is, ultimately, only a *description* of the planned analysis, rather than the actual analytical code, which constitutes the analysis itself. Ideally, having both the plan and the code would provide detail of the planned analyses, the rationale behind these decisions, and how this was implemented, giving a fully transparent picture. Some software applications or languages (e.g., R, Python) have multiple versions of analysis procedures (i.e., packages) for a given method and each package may differ slightly in how the method is implemented – each equally valid, but written differently due to peculiarities of that particular programmer. For example, the R packages

{lme4} (via either the lmer() or glmer() functions), {nlme}, and {lmm} are all capable of implementing a valid linear mixed effects model, with each differing slightly in how the parameters are estimated. Another example is with Bayesian analyses: using different sampling algorithms to extract posterior distributions from the same analysis model (e.g., Gibbs vs. Hamiltonian sampling). Multiple imputation, which handles how to impute (estimate) plausible values for missing observations in a dataset, is another increasingly common example of a complex analytical method, rarely described in detail in study manuscripts, which has potential for substantive differences in analysis results depending on the (many) specific analytical decisions taken (see Table 4 for examples of potential sources for variation relating to the model specification and code). An R script showing these examples implemented in code are hosted at this [Open Science Framework Repository](). These seemingly minor differences in analysis implementations can result in differences in processing speed, how estimates are calculated, the available options that can be included in the procedure, such that simply describing in the SAP that 'mixed effects models' or 'multiple imputation by chained equations' are to be used is insufficient to exactly replicate the analysis or give readers enough detail to understand the model specification. Describing only which procedure is planned on being used (as is typical in a SAP) or actually used (in the results paper) creates uncertainty about exactly which packages were used and how this was implemented in code. Factors such as how these models were programmed, the specific order in which they were implemented, and the specific package versions may also vary depending on the particular programmer's experiences, personal preferences, and analytical skill. This results in many sources of potential ambiguity that are not able to be scrutinised without direct access to this code, but all of which may have a meaningful impact on the results. Lastly, there may also be inadvertent errors in the code that can have a material impact on the results and associated inferences. Only by directly seeing the analytical code, and appropriate documentation of metadata and the 'environment' (operating system, software versions, packages, and similar), are readers able to fully scrutinise the decisions made by a research team, and decide whether their operationalisation of the decisions are correct and appropriate.

Investigations that require larger, more complicated data sources, or more complex methodology can often be analysed using different, seemingly valid, analytical strategies (relative to something more straightforward and simple ones where only a few analytical strategies are possible). Because complexity inherently creates more possible different ways to 'tackle' a research question, there is an increased need to be able to scrutinise these strategies. However, paradoxically, details about the decisions made in a more complex and convoluted analysis are perhaps *less* likely to be included in the manuscript or SAP due to the sheer number of analytical decisions required, or perhaps because the decisions are considered by the authors to be too specific, niche, or irrelevant to be included. Even when methods that were used to carry out the analyses are described, the detail is often insufficient to exactly reproduce the analyses, with crucial details often omitted. Some examples of these critical but under-reported decisions are how categorical variables are coded, a complete description of how missing data is handled, full details on the training of machine learning and artificial intelligence models, the specific algorithm used, and seed values for pseudo-random number generation (**Table 1**).

The ability to review analytic code can improve trustworthiness, aid reproducibility (using code or data to reproduce results)[7], and increase transparency[8,9]. Sharing code also facilitates other research teams to build on existing analyses - if some study timepoints were omitted or some assumptions about analyses made, (for example how a longitudinal trajectory was modelled in a study), another team can reproduce the coding environment, take the extant code, carry out additional analyses, and then produce additional supplementary findings from an existing paper rapidly (provided they also have the study dataset). As a result, many supporters of open research practices have argued for the inclusion of analytic code in submissions[10–13]. Furthermore, there is a trend toward making inclusion of analytic code a standard practice (even when the study dataset cannot be shared), and some journals are in the process of - or have already begun - implementing mechanisms for the inclusion of analysis code[14–16], while at otherjournals this requirement is being developed[17]. We argue that this is still favourable to not sharing code at all, as code sharing not onlyincreases transparency and accountability, but also has educational value and code might be reused with synthetic or similar datasets. In these cases, this would enable quick adaptation rather than having to code from scratch.

While the need for scrutiny is therefore warranted, peer review and public post-publication review of data analyses is difficult due to the lack of shared analytic code[18].

Moreover, this deficit also negatively impacts the opportunity to learn from another scientific team's coding decisions and methodologic implementation. Code sharing can further contribute to a more inclusive software environment for researchers with limited resources, enabling them to access content and code that would usually require payment of fees or licensing costs. Whilst this might run counter to commercial interests and the monetisation of intellectual property, it is important to remember that most research is funded by taxpayers or charitable bodies, and thus constitutes a public interest task that should produce the most accessible outputs. Sharing code from tasks regularly done by many research teams (e.g., scoring a validated open-source general health questionnaire) may also increase efficiency - others can use that same code and save unnecessary duplication. These benefits and the need for code availability have been recognised[4,9,10,12,14–16,19], and most recently the BMJ became the first medical journal requiring code availability[20]. Appropriate licensing of code and software is therefore imperative to enable free and open access to code and software, whilst safeguarding commercial interests and unintended use.

Here we outline some basic principles for analytic code sharing, as we acknowledge that while there is a need to increase sharing of actual code (rather than mere descriptions of analyses), simply uploading code as developed by a research team with little preparation, explanation, and justification may provide little additional value. As discussed, analysts increasingly write code, and likely have not had formal training in good coding practices: either because they have not had exposure to such training, or because - aside of those used in small independent research groups or commercial software developers - standardised frameworks for good coding practice do not exist in health research. We therefore outline some high-level principles of good practice in code sharing to help foster not only the sharing of analytical code, but sharing of *good quality*, *scrutable* code that is beneficial to the reader and researchers more generally. By applying these concepts in their own work, researchers can improve the transparency and reproducibility of their research, rather than simply uploading poor quality code to satisfy a journal's data sharing requirements. Whilst we appreciate that health research often involves barriers to fully reproducible outputs, such as working with sensitive, proprietary or restricted data that cannot be shared, these barriers generally do not apply to analytical code – as opposed to source code - and we therefore want to ensure that outputs are adhering to the *FAIR Data Principles*[21,22].

**Principles of Code Sharing**

A key driver behind code sharing is *transparency* and *reproducibility*[7,23–25]: documenting both how the analysis was done, but importantly, why it was done in a particular way. While the code itself discloses exactly what was done in the analysis (the 'how'), the rationale explaining *why* these decisions were made requires a more in-depth description of what each section of the code is doing (particularly if that section of code is doing something that might be considered by a reader to be idiosyncratic) - typically done in the form of *code comments* written by the analyst. Generally, the more exposition the better, as long as the additional text does not impact clarity. Good code sharing practice can be summarised with a few core principles that are broadly applicable, and can be applied in every case of research publication. While FAIR provides guidelines for open *data*, the same principles can be applied and adapted to analysis code. Hence, we retain the four core FAIR principles that code, like data, should be **Findable, Accessible, Interoperable, and Reusable**, and the specifics of how these principles apply to analysis code are outlined below. A summary checklist of this information that can be used as a 'handy guide' is available in Table 2.

*F: Code should be findable.*

*Findability* (also known as 'discoverability') refers to a research team formatting and storing analysis code in such a way that allows other researchers to easily locate the code directly from the article or upon an internet search. Ideally, code will be persistently associated with the published manuscript as an addendum or supplement maintained by the journal much like supplemental files are currently indexed. The code can be contained in a document file (e.g. copied and pasted into a word-processor file format *.doc, *.docx, or *.pdf) or more preferably in a plain-text file (i.e., *.txt) or a software specific file format (e.g., *.Qmd, *.R, *.py, *.do, *.sas). While manuscripts are increasingly given persistent identifiers (digital object identifiers, DOIs), code is often included in journals as an additional supplement which can often be subject to 'link rot' (i.e., hyperlinks no longer working)[26]. Ideally, rather than being retained on a journal website, code can also be stored in openly accessible repositories (e.g., GitHub, GitLab), some of which produce their own permanent links or identifiers, to help ensure that the code is persistently available for readers (e.g, Zenodo, Figshare). Many of these repositories also have version control/history to allow tracking of changes and updates on any code edits. Another aspect of persistent identifiers is their discoverability by computers, making them human and machine readable and enhancing findability whilst making them easy to resolve on the web.

While we have no position regarding the 'best' location to keep code, some examples of well-known and maintained (as of publication) open access repositories are included in **Table 3**. Open access repositories are appealing because they are built with code sharing in mind. Thus, even if journals have a method to store code with the original paper, we consider open access code repositories to be preferable. The importance here is that when looking for the code from a manuscript it is 1) easy to find and confirm that the code is meant specifically for the manuscript's analysis and 2) that it is available in perpetuity. We would discourage analysts from storing code on personal websites, for example.

Much of what makes a manuscript and its code more discoverable relates to the quality of the descriptive *metadata* included with the code or manuscript rather than any of the document content itself. Keywords and similar information that help flag the topics, such as disease areas, authors, related laboratories, trial identifiers, relevant to the code should be included. For example, code files can include a 'header' section at the start of the file - explaining the title, purpose, date of creation, date modified, authors - or accompanying information, such as a 'readme file' - added to the repository when a document is submitted or stored which contains information on *what* the code or software is performing when executed, *how* to execute the code script, the necessary data and environments to do so, and the licenses this code operates under) to help contextualise a code file or manuscript and clearly outline the steps necessary to run the code. This can make the files more likely to be picked up by search queries, and also helps authors identify similar files which are also relevant. This is particularly true if multiple studies form part of an interlinked series.

*A: Code should be accessible*
Analysis code is typically written in plain-text format, which is designed to be accessible. hile different code files may use different file extensions depending on the software, oftentimes they are human-readable and 'openable' in plain-text editing software (e.g. Notepad, Visual Studio Code, Emacs). However, merely listing lines of code with no explanation may be difficult to understand due to the complexity of commands needed for a particular analysis or to achieve efficiency. Thus, analysis code should be well documented within the file itself[25]; this is where header, Readme documentation, and commenting play a crucial role.

All files should include associated header information and possibly a Readme file (link to example here hosted on OSF repository) to document all relevant details of the analysis file including those who coded the analysis, the names of any datasets used, and a summary of what each file contains, dependencies, and prerequisites for execution (i.e., any other extraneous code scripts/software/data that is required to execute the code successfully without errors)[25]. Furthermore, code will be adequately (and consistently) commented so that readers can easily follow exactly what is being accomplished with each code block (or at least what the analyst believes is being accomplished), and *why*. This will mean that each coding step is ideally well described, every new command is referenced in the context of the entire analysis, and links to the methods section of the main paper and supplement are made. To quote the late Professor Doug Altman, "Readers should not have to infer what was probably done; they should be told explicitly"[27]. For instance, something as simple as loading all packages used in the analysis and their versions in R or Python should be coded, with a comment or note that explains *why* the package is being included so that the reader understands generally what it used for in the analysis.

All variables should be referred to in human readable forms using 'labels' where software allows. An example is referring to a variable named `wt81` with the label 'weight in pounds in 1981' instead of just using the short variable name alone without describing what this variable is. This will increase clarity and readability. **Box 1** illustrates examples of poorly, reasonably, and well annotated code. Programming style guides, which provide an explicit set of conventions for formatting, naming, and styling code, can facilitate coding in an organised and consistent manner. A coder's development environment may also include the ability to 'lint' code – where the software automatically spots and highlights inconsistencies with a coding style to make coding more consistent and in-keeping with a style guide (much akin to a spellchecker in word processing software). Examples of software that has this ability includes, but is not limited to, Visual Studio Code, RStudio, Positron, and Sublime Text. Producing a 'codebook' - a separate document listing variable names, labels, their possible values (if categorical), and how missing values are coded (if relevant) for a given dataset - are particularly useful with larger datasets that contain many variables, but can take additional effort and are not essential if variable definitions are clear and human-readable in the script.

Coding, in practice, is often organised in a project folder structure rather than as isolated code files. As well as clearly denoting and organising the code content, variable names, and metadata, when sharing a structured folder which might contain multiple code files, documentation, datasets, it is important to maintain a logical, human-readable folder structure that is easily understood. Examples of organised folder structures are included in **Figure 1**.

*Licensing*

An important aspect of code accessibility is the issue of appropriate licenses for sharing, reusing, and modifying of code. These licenses can be added to the repositories and cover commercial and non-commercial purposes. Commonly attributed licenses, and tools to choose the most appropriate one, can be found online (choosealicense.com, creativecommons.com). We suggest that it is best practice to include a license when sharing code so that readers have clarity on how the authors intend their code to be used, protect their intellectual property, safeguard commercial interests, and gain appropriate credit. This includes acknowledgment of others' work through proper citation.

**I: Code should be interoperable (implementable).**

Beyond its accessibility, code will need to be implementable by people distinct from the original authors. By *interoperable* we mean that the code is readable by a computer and will work in its entirety when run by a different user, on the original analysis dataset, or another dataset which has the same format and structure as the original analysis dataset - in short, the code will "work" when used in the proper context, regardless of who uses it, or where it is used. This includes any code that produces *all aspects* of the analysis output: results, and any prior steps used to produce those results, as well as any figures, and tables. The guiding principle should be that the reader should be able to make a through-line from any result included in a manuscript, right back to the dataset from which it was derived, allowing them to fully scrutinise every result. This means that it contains the entirety of the analysis (including any sensitivity analyses), documenting all steps for data preparation and cleaning. However, it does not guarantee that the results produced are identical to that of the original manuscript. Any expected warnings or errors should be documented and explained; no unexpected errors should arise on repeated code execution. The code should also include any analysis presented in a manuscript's supplemental content.

*Software versioning*

Software versioning can become an issue, and it is important to ensure that the code files contain information about the software used to create the file and its version, so that, should a reader wish to use the code file, they know (and which additional packages or extensions are used if using software which requires additional extensions). An example of this would be using the `sessionInfo()` command in R to obtain the specific R version used, the operating system and any attached packages and their corresponding versions. An example of this output is included in **Box 1c**.

### R: Code should be reusable.

In order to facilitate reusability , all unique components of the analysis, such as data sources, parameters, and processing steps, should be clearly defined and embedded within the code, enabling others to use, adapt or extend it to different contexts with minimal effort. This includes, but is not limited to, any seeds used, e.g., simulating data and stochastic processes, all names and versions of software, software add-ons (i.e., packages that need to be added to extend the functionality of base software) and date of analysis should be explicitly listed, and either the original (raw) data, simulated/synthetic data or metadata. Where data sharing is not possible, explaining the rationale for *why* the dataset was not shared and possible routes to access it should be included, avoiding platitudes like the "data are available upon reasonable request" without adding more context. Providing all this information is particularly important for software that makes extensive use of add-ons (e.g., when using R or Python). For software like SAS, this might require inclusion of macros that were used in addition to references to the source. It should also be confirmed that the extension/macro/add-on can be run independently on different computers than those that completed the analysis. Reusability is perhaps the most challenging aspect of code sharing, as it is complex, and requires additional technical computing knowledge not typically taught alongside statistical, analytical, or clinical training. This may be difficult to achieve, but we suggest aiming for the highest possible level of reuse, to facilitate computational reproducibility[28]. Many technical factors can result in small variations in the estimates produced by identical code, including but not limited to: the version of the software (and extensions/packages/add-ons) used; the operating system on which the code is run; the seed used to initiate any random number generators; the algorithm of the random number generator used; other operating system-level software present on the machine used; networking, access rights, hardware requirements, and similar issues of the machine used to run the analytical code.

We suggest that the best way to overcome the complexity is transparency through applying the FAIR criteria outlined above and including as much information about how (and why) the analysis was carried out as possible.
This can be achieved by using package and environment management tools (e.g., Conda, Poetry or uv) or by *containerising* an application, which packs software and all necessary packages and dependencies into a virtual 'container' so it can run consistently across different environments (e.g., Docker). Another helpful tool specifically focussed on sharing and executing code is Binder[29]. Fully 'containerised' code that can be shared between machines and produces the exact estimates as they were on the original machine is the ideal solution, however we accept that containerising analysis code can be technically demanding, with limited guidance and training available aimed at clinical research.

**Code review and double coding**
Code review is an important tool to help researchers ensure that analysis code is correct, and the decisions made in the code is consistent with the description of the analysis in the manuscript. This involves a separate analyst going through the analysis code line-by-line to make sure the description and decisions are clear, expected results are reproduced and any errors or ambiguity is identified and rectified. While this process can be time consuming and not without cost, it helps to minimise mistakes and errors. Because it involves a separate analyst, this process can also help researchers ensure the code can run on other computers and simulates what might be expected if an external investigator were to run the code.

A step up from code review is the process of double coding where an analysis is performed twice, by two different analysts independently. This approach can be further bolstered by coding analyses in ≥ 2 different software types (e.g., SAS and R, R and Python, SAS and Stata). In this way the research team can be reassured that the analysis comes to the same (or quite similar) results using potentially different approaches. This also has the benefit of allowing the code to be shared in more than one software in the eventual publication, broadening the reach of the analytic methods. While this practice may be considered resource intensive, it is common practice in some clinical trials units and regulators (e.g., the US Food &

Drug Administration) to at least double code the primary analyses of a clinical trial due to its potential for significant impact on clinical practice.

Whichever method was chosen, most funding bodies now allow costs to be charged to these tasks to ensure code and data sharing are done properly and sustainably.

**Perceived barriers to sharing code**

Some concerns have been raised by analysts when considering whether to include analytic code. First, there might be apprehension on the part of the authors sharing code because they feel it is not 'optimised', may be poorly written, and may contain errors, and thus might result in embarrassment[10]. While there is no 'silver bullet' to alleviate this concern, we stress that the benefits of code sharing and the risks posed by not sharing code greatly outweigh these concerns: we feel that there is a duty for research to uphold scientific integrity when their code underpins clinical findings that may result in changes to clinical practice and has subsequent clinical and financial impact on health and care. There are strategies that can mitigate anxiety around sharing code[13], and a recent blog post outlines a personal experience of code sharing that we hope can help allay fears anxieties around 'showing your work'[30].

A second concern would be a claim of proprietary software and code use or variable names that might be considered proprietary by data brokers or owners (for example, if using an outcome in analysis such as step count, which is derived from sensor data collected on a wearable device, and derived using a proprietary algorithm). In the case of proprietary variable names, the code can simply use different names for the same variable. In the case of proprietary code or software, the situation is more complex. However, we would advocate including a notation of which code in the analysis is proprietary, the ownership of the code, the purpose of the code (in case the original code may not be available, to help guide understanding and potential workarounds), and mechanisms to gain access to the code (if possible). In this way it would allow any subsequent researchers to better understand the restrictions around the analysis code and how to obtain it for themselves. While data protection legislation such as the Privacy Act 1974 or the General Data Protection Regulation (GDPR) is relevant to the study datasets and appropriate protections should exist around identifiable, sensitive and personal data, analytical code does not regularly contain identifiable information or data. Where this is a concern, e.g., inferring identifiable information from machine learning weights or small counts, adequate explanations are still important and code review also applies to these restricted environments considering that more and more research is conducted in Trusted Research Environments which require code cleaning and anonymisation before exiting virtual machines. The guiding principle of code sharing should be around maximising transparency and scrutability, and we find it difficult to reasonably justify the specific withholding of analytical code on these grounds outside of commercial/patent protections.

A third concern may be with respect to peer review: that reviewers of code need to have the technical skill and the resources to be able to review and scrutinise code with sufficient ability and depth to understand whether the analyses were undertaken correctly and appropriately. No reviewer is expected to be an expert of all aspects of a manuscript, and we maintain that sharing of code to maximise transparency should be the *default* position to allow scrutiny at the time of review and in future. Indeed, emerging and revised guidelines are beginning to introduce code sharing in their recommendations (e.g., CONSORT 2025[31,32]; TRIPOD+AI[33]), encouraging this to be standard practice.

**Advice regarding reviewing papers with analytic code**

We take the view that code submission is for *long term* transparency and for utility and verification after publication and meant for much more than verification solely during any peer review prior to publication of a study's results. We consider the analysis code as an essential part of research conduct, akin to the protocol, analysis plan, and research findings, and therefore feel it should be present alongside other documentation in perpetuity to allow proper scrutiny and understanding well after initial publication. Nevertheless, if reviewers do feel qualified to review code in addition to the text, they should do so. Moreover, authors should ideally consider code review as 'fair game' for consideration as is the remainder of the manuscript or supplementary materials–it is essentially a more in-depth review of a study's methods. At a minimum, code review by a paper reviewer should check to make sure it conforms to the FAIR principles and consider documentation and transparency as outlined above (see **Table 2** for some considerations). More in depth review would involve comparing the description of the methods in the paper to make sure the code generally agrees with what is claimed in the described methodology. Additionally, reviewers can consider running the code independently to make sure results are the same as the authors. If discrepancies are identified,

OS, software, and package versions should be offered to the authors in the review to troubleshoot any bugs.

Code review for a journal can be daunting even for qualified manuscript referees; it can be burdensome on top of the already time-consuming review process. Further, reviewers might not be familiar with particular software, methods, etc. Instead, it is important to verify any unclear claims in a paper or for able and interested referees. Thus, reviewers should examine pieces of the analysis as needed or as desired.

**Future developments that foster transparency**

More recently, analysis code may also be *directly woven* into study manuscripts using 'literate programming' approaches (e.g., Jupyter Notebooks; Quarto; Dyndoc[34–36]), making the manuscript and analysis code inseparable. The implementation of these methods is outside the scope of this paper, but it is clear how these approaches can help facilitate greater transparency and reproducibility still than simply sharing the code as a discrete file, and the authors highlight this as a model of best practice for maximising openness and transparency with respect to analyses.

**CONCLUSION**

As research becomes increasingly complex and computational, the ability to share, understand, and build upon analytical code is essential for transparency and good scientific practice. By adopting the FAIR principles, applying them to code-sharing, and thus embedding good coding practices into the research lifecycle, we can shift the culture from code being a by-product of research to a shared, valuable output.

Whilst there might be prejudices towards code-sharing, primarily around it being perceived as time-intensive and not always delivering the hoped-for value, we argue that once the necessary infrastructure and practices are in place, we can shift the perception of it from being an optional add-on, to a core component of rigorous and responsible research. With the right support, the benefits of code sharing by far outweigh the perceived burdens, making our research more open, transparent, and impactful.
**An Open Science Framework repository containing example code adhering to these principles is available [here](here).**

**Summary Points:**

- Modern quantitative research, including - but not limited to - health care, psychological, economic, epidemiological, and biostatistics research relies on coding languages in order to execute analyses, from data cleaning and harmonisation to the fitting or building of analytical models (e.g. statistical, machine-learning, etc.) and the derivation of estimates, uncertainty quantification and predictions.
- Analysts can therefore be considered to contribute toward software development and methodology implementation, creating detailed analytical code, the content for which is not standardised or controlled. Analysts may adhere to myriad patchwork, informal, and incomplete standards for coding practice and quality, to varying degrees.
- Analytical decisions and implementations have wide and important ramifications for the results of research analyses. However, many analytic decisions are not described in the methods portion of manuscripts.
- Sharing analysis code can make analytic decisions more transparent, but the process of sharing analysis code and best practices are not widely available.
- Here we describe important aspects to encourage and improve the quality of sharing analysis code when submitting a manuscript as a pre-print, for peer review, or for publication.

**References**

1 Gelman A, Loken E. The Statistical Crisis in Science. *Am Sci*. 2014;102:460. doi: 10.1511/2014.111.460

2 Gelman A, Loken E. The garden of forking paths: Why multiple comparisons can be a problem, even when there is no "fishing expedition" or "p-hacking" and the research hypothesis was posited ahead of time.

3 Schwab S, Held L. Statistical Programming: Small Mistakes, Big Impacts. *Significance*. 2021;18:6–7. doi: 10.1111/1740-9713.01522

4 Huebner M, Vach W, Le Cessie S, *et al.* Hidden analyses: a review of reporting practice and recommendations for more transparent reporting of initial data analyses. *BMC Med Res Methodol*. 2020;20:61. doi: 10.1186/s12874-020-00942-y

5 Huntington-Klein N, Arenas A, Beam E, *et al.* The influence of hidden researcher decisions in applied microeconomics. *Econ Inq*. 2021;59:944–60. doi: 10.1111/ecin.12992

6 Silberzahn R, Uhlmann EL, Martin DP, *et al.* Many Analysts, One Data Set: Making Transparent How Variations in Analytic Choices Affect Results. *Adv Methods Pract Psychol Sci*. 2018;1:337–56. doi: 10.1177/2515245917747646

7 Committee on Reproducibility and Replicability in Science, Board on Behavioral, Cognitive, and Sensory Sciences, Committee on National Statistics, *et al. Reproducibility and Replicability in Science*. Washington, D.C.: National Academies Press 2019.

8 Brown AW, Kaiser KA, Allison DB. Issues with data and analyses: Errors, underlying themes, and potential solutions. *Proc Natl Acad Sci*. 2018;115:2563–70. doi: 10.1073/pnas.1708279115

9 Nallamothu BK. Trust, But Verify. *Circ Cardiovasc Qual Outcomes*. 2019;12:e005942. doi: 10.1161/CIRCOUTCOMES.119.005942

10 Gomes DGE, Pottier P, Crystal-Ornelas R, *et al.* Why don't we share data and code? Perceived barriers and benefits to public archiving practices. *Proc R Soc B Biol Sci*. 2022;289:20221113. doi: 10.1098/rspb.2022.1113

11 Munafò MR, Nosek BA, Bishop DVM, *et al.* A manifesto for reproducible science. *Nat Hum Behav*. 2017;1:0021. doi: 10.1038/s41562-016-0021

12 Goldacre B, Morton CE, DeVito NJ. Why researchers should share their analytic code. *BMJ*. 2019;l6365. doi: 10.1136/bmj.l6365

13 Software Sustainability Institute. Top ten reasons to not share your code (and why you should anyway). 2013. https://www.software.ac.uk/blog/top-ten-reasons-not-share-your-code-and-why-you-should-anyway (accessed 14 May 2025) 14 Nature Journal. Nature Portfolio Editorial Policies, Reporting Standards and Availability of Data: Availability and peer review of computer code and algorithm. https://www.nature.com/nature-portfolio/editorial-policies/reporting-standards#availability-of-data (accessed 14 May 2025)

15 European Urology Journal. Guide for Authors: Submission of Statistical Code. https://www.elsevier.com/journals/european-urology/0302-2838/guide-for-authors

16 PLoS Medicine. Materials, Software and Code Sharing. https://journals.plos.org/plosmedicine/s/materials-software-and-code-sharing

17 Abbasi K. A commitment to act on data sharing. *BMJ*. 2023;p1609. doi: 10.1136/bmj.p1609

18 Peng RD. Reproducible Research in Computational Science. *Science*. 2011;334:1226–7. doi: 10.1126/science.1213847

19 Assel M, Vickers AJ. Statistical Code for Clinical Research Papers in a High-Impact Specialist Medical Journal. *Ann Intern Med*. 2018;168:832–3. doi: 10.7326/M17-2863

20 Loder E, Macdonald H, Bloom T, *et al.* Mandatory data and code sharing for research published by *The BMJ*. *BMJ*. 2024;q324. doi: 10.1136/bmj.q324

21 Barker M, Chue Hong NP, Katz DS, *et al.* Introducing the FAIR Principles for research software. *Sci Data*. 2022;9:622. doi: 10.1038/s41597-022-01710-x

22 Wilkinson MD, Dumontier M, Aalbersberg IjJ, *et al.* The FAIR Guiding Principles for scientific data management and stewardship. *Sci Data*. 2016;3:160018. doi: 10.1038/sdata.2016.18

23 Wilson G, Bryan J, Cranston K, *et al.* Good enough practices in scientific computing. *PLOS Comput Biol*. 2017;13:e1005510. doi: 10.1371/journal.pcbi.1005510

24 Lee G, Bacon S, Bush I, *et al.* Barely sufficient practices in scientific computing. *Patterns*. 2021;2:100206. doi: 10.1016/j.patter.2021.100206

25 Hochheimer CJ, Bosma GN, Gunn-Sandell L, *et al.* Reproducible research practices: A tool for effective and efficient leadership in collaborative statistics. *Stat*. 2024;13:e653. doi: 10.1002/sta4.653

26 Pew Research Center, Chapekis A, Bestvater S, *et al.* When Online Content Disappears. Pew Research Center 2024.

27 Altman DG. Better reporting of randomised controlled trials: the CONSORT statement. *BMJ*. 1996;313:570–1. doi: 10.1136/bmj.313.7057.570

28 Freire J, Chirigati F. Provenance and the Different Flavors of Computational Reproducibility. *IEEE Data Eng Bull*. 2018;41:15–26.

29 Jupyter P, Bussonnier M, Forde J, *et al.* Binder 2.0 - Reproducible, interactive, sharable environments for science at scale. Austin, Texas 2018:113–20.

30　Dahly D. Open science is really scary y'all. Life Pain Espec. Your Data. 2020. https://statsepi.substack.com/p/open-science-is-really-scary-yall (accessed 14 May 2025)

31　Hopewell S, Chan A-W, Collins GS, *et al.* CONSORT 2025 statement: updated guideline for reporting randomised trials. *BMJ*. 2025;389:e081123. doi: 10.1136/bmj-2024-081123

32　Moher D, Collins G, Hoffmann T, *et al.* Reporting on data sharing: executive position of the EQUATOR Network. *BMJ*. 2024;e079694. doi: 10.1136/bmj-2024-079694

33　Collins GS, Moons KGM, Dhiman P, *et al.* TRIPOD+AI statement: updated guidance for reporting clinical prediction models that use regression or machine learning methods. *BMJ*. 2024;e078378. doi: 10.1136/bmj-2023-078378

34　Allaire JJ, Teague C, Xie Y, *et al.* Quarto. 2022.

35　Kluyver Thomas, Ragan-Kelley Benjamin, Pérez Fernando, *et al.* Jupyter Notebooks - a publishing format for reproducible computational workflows. *Positioning and Power in Academic Publishing: Players, Agents and Agendas*. IOS Press 2016.

36　StataCorp. *Stata 19 Base Reference Manual: dyndoc*. College Station, TX: Stata Press 2025.

37　Smith GB, Redfern OC, Pimentel MA, *et al.* The National Early Warning Score 2 (NEWS2). *Clin Med*. 2019;19:260. doi: 10.7861/clinmedicine.19-3-260

**Table 1: A checklist for code-sharing considering the FAIR principles, documentation, and transparency**

| | |
|---|---|
| FINDABILITY | ❑ Code is stored in a public, persistent repository (e.g., GitHub, Zenodo, OSF)<br>❑ Repository or code supplement are linked directly from the manuscript to the project page<br>❑ Persistent identifier (e.g., DOI) is provided |
| ACCESSIBILITY | ❑ Code is in plain-text, human-readable format (e.g., .R, .py, .do, .txt, etc.)<br>❑ A README file is included, describing the code's purpose, usage, and dependencies<br>❑ All files have a header with the title, author(s), data, and version information<br>❑ Code is adequately commented to explain each step and decision |
| INTEROPERABILITY/ IMPLEMENTABILITY | ❑ All required software, package versions, and operating system details are specified (e.g., via `sessionInfo()` in R or a YAML file)<br>❑ All data preparation, cleaning, and analysis steps are included in the code<br>❑ Any expected warnings or errors are documented<br>❑ Code runs from start to finish without modification on a new machine (given the same data structure) |
| REUSABILITY | ❑ All unique parameters, seeds, and processing steps are explicitly defined in the code<br>❑ Code is modular and structured for adaptation to new datasets or contexts<br>❑ Licensing information is included, specifying terms of use and citation<br>❑ If data cannot be shared, code to generate simulated/synthetic data is provided or an explanation why code cannot be shared |
| DOCUMENTATION AND METADATA | ❑ Variable names are descriptive and, where possible, labelled<br>❑ A codebook or glossary is provided for variables and key terms<br>❑ Folder structure is logical and documented<br>❑ Instructions for reproducing figures and tables are included |
| TRANSPARENCY AND REVIEW | ❑ Analytical decisions and rationale are documented (in comments or supplementary material)<br>❑ Code has been peer-reviewed (e.g., pull requests or journal reviewers) or double-coded where feasible |

|  | ❏ Any proprietary code or dependencies are clearly identified, with instructions for access or alternatives |
| --- | --- |

**Table 2: Examples of specific analytical processes that are susceptible to material effects on analysis model estimates if not defined clearly**

| Table 1 | | |
| --- | --- | --- |
| **Examples** | **Description/Example** | **Problem** |
| Multiple Imputation (MI) | Methods for handling missing data in a clinical trial analysis such as multiple imputation (MI) may be described in the manuscript briefly, but MI is a model in its own right, separate and distinct from the inferential analysis model (e.g. a mixed effects model to decide whether two treatment groups differ). However, decisions such as which modelling variables to include and whether to incorporate auxiliary information, etc., are omitted. | Decisions can have a material effect on the bias, precision, and underlying assumptions of subsequent estimates in the main analysis model (see table 4 for examples). |
| Variable Definition | Whether categorical variables (for example, the modified early warning score, NEWS2[37]) are treated as distinct categories, ordered | How variables are treated in the analyses is also a source of potential differences between analyses and may also affect the results and |
| | 'ranks' (ordinal variables), or as a continuous scale. | subsequent interpretation of findings. |
| Machine learning and other artificial intelligence augmented analyses | These are complex, data-driven, algorithm-led analyses which can produce different outputs given different starting parameters in the code and/or different training datasets, even when the same method and packages are used on the same analysis dataset. Even describing what methods were used in the manuscript (which, as have outlined, does not occur fully or regularly in many cases) is not sufficient to exactly reproduce the analyses without these additional parameters being defined: hence the need to see the *actual code* underlying these analyses. | Without all the initial starting conditions the algorithms may produce starkly different outputs. Moreover, the integrity and validity of the modelling may be hard to verify. |

**Table 3: Examples of Repositories for Storing Analytical Code**

| Repository Name | Funding | Permanence? | Notes |
|---|---|---|---|
| Github | Private | Likely high. Est. 2008. | Web repository for local git repositories that can sync local files to remote cloud locations and shared publicly. |
| Gitlab | Private | Likely high, Est 2018 | Same as GitHub |
| Bitbucket | Open Source, privately owned | Likely high, Est 2008 | Same as GitHub |
| Open Science Framework(OSF) | Open source, funded by philanthropic ventures | Moderate, Est 2013 but depends on ongoing funding | Not primarily for software. Hosts multiple types of files including preprints and protocols. |
| Zenodo | Open source, funded and operated by CERN | Moderate-High, Est 2013 but depends on ongoing funding | Similar to OSF, but with greater integration with Github |
| Octopus.ac | Open source, managed in partnership by Jisc and Octopus | Likely high, funded by UK government, Est 2022 | Relatively new platform, intention to allow registration of manuscripts, code, protocols, but also ideas for studies. Aimed to link and record all elements of the research cycle. |

| | Publishing CIC, funded by UKRI | | |
|---|---|---|---|

**Table 4: Potential Sources of Variation when Implementing a Multiple Imputation by Chained Equations (MICE) Model, Not Typically Described in Manuscripts, with Authors' Deemed Potential for Impact on Results**

| Element of Imputation Model | Implementation Step | Risk Potential for Substantive Impact on Results |
|---|---|---|
| Model specification decisions | Matching/imputation algorithm used (e.g. predictive mean matching; multivariate normal) | High |
| | Different choices of auxiliary variables (which non-analyses variables are included in the model to improve model fit) | High |
| | Handling of passive (derived/calculated variables), whether:<br>- Just another variable (JAV)<br>- Passive imputation<br>- etc. | High |
| | Whether imputed values are constrained to the range of the included outcome(s) | Medium/high |
| | Whether, and how, interaction effects are included in the model | High |
| Computing/coding decisions | Software package used (e.g. R, SAS, Stata) | Medium/high |
| | Different MICE packages (e.g. {mice}; {mi}; {micemd}; etc. in R) | Medium/high |
| | Seed used to initiate random number generators | Low |
| | Number of 'burn-in' iterations before datasets are saved | Low |
| | Tolerance parameters for imputation model convergence | Potentially high |
| | Number of imputed datasets retained | Medium |
| | Storage of imputed datasets (fully wide, fully long, etc.) | Low |

**Box 1**

> A. **Inadequate, poorly formatted and annotated code**: Code is bunched without spaces to delineate different analysis steps. There is no annotation describing what is being done or why. This would be difficult to follow, particularly for long analysis code

```
library(here)
library("readxl")
nhefs <- read_excel(here("data", "NHEFS.xls"))
nhefs$cens <- ifelse(is.na(nhefs$wt82), 1, 0)
nhefs.nmv <-
  nhefs[which(!is.na(nhefs$wt82)),]
lm(wt82_71 ~ qsmk, data = nhefs.nmv)
```

> B. **Better formatted and annotated code**: Separation between analysis into logical partitions aids in readability. There is some annotation, but it is not extensive. This would be easier to read and recreate compared to the above, but still not ideal for someone coming into the project fresh or after a long hiatus. We would consider this 'adequate', but not ideal.

```
library(here)

# install.packages("readxl") # install package if required
library("readxl")

nhefs <- read_excel(here("data", "NHEFS.xls"))
nhefs$cens <- ifelse(is.na(nhefs$wt82), 1, 0)

# provisionally ignore subjects with missing values for weight in 1982
nhefs.nmv <-
  nhefs[which(!is.na(nhefs$wt82)),]

lm(wt82_71 ~ qsmk, data = nhefs.nmv)
```

> C. **Idealized formatted and annotated code**: Every step is annotated and decision highlighted. Session information, which includes package versions, is included at the end of the code. Code commented in this way is much more easily understandable to a non-technical reader than examples A or B.

```
#load package 'here' that looks for files in current directory
library(here)
```

```
# install.packages("readxl") # install package if required
library("readxl") #load 'readxl' to read data file

#read excel data file, filename='NHEFS.xls' in folder /data/ and store it in
#object nhefs.
nhefs <- read_excel(here("data", "NHEFS.xls"))

#create a missing indicator where if variable 'weight in 82' is missing variable
#'cens' is 1, and zero when not missing.
nhefs$cens <- ifelse(is.na(nhefs$wt82), 1, 0)

# provisionally ignore subjects with missing values for weight in 82
nhefs.nmv <-nhefs[which(!is.na(nhefs$wt82)),]

#linear model regression change in weight between 1971 and 1982 on quantity
#smoked using data with missing values with weight in 82 excluded.
lm(wt82_71 ~ qsmk, data = nhefs.nmv)

#Session Info:
R version 4.3.1 (2023-06-16)
Platform: x86_64-apple-darwin20 (64-bit)
Running under: macOS Sonoma 14.3.1

attached base packages:
[1] stats     graphics  grDevices utils     datasets  methods   base

other attached packages:
[1] readxl_1.4.2 here_1.0.1
```

**Figure 1: Example project folder structure for a large, complex project (a) and small standalone project (b)**

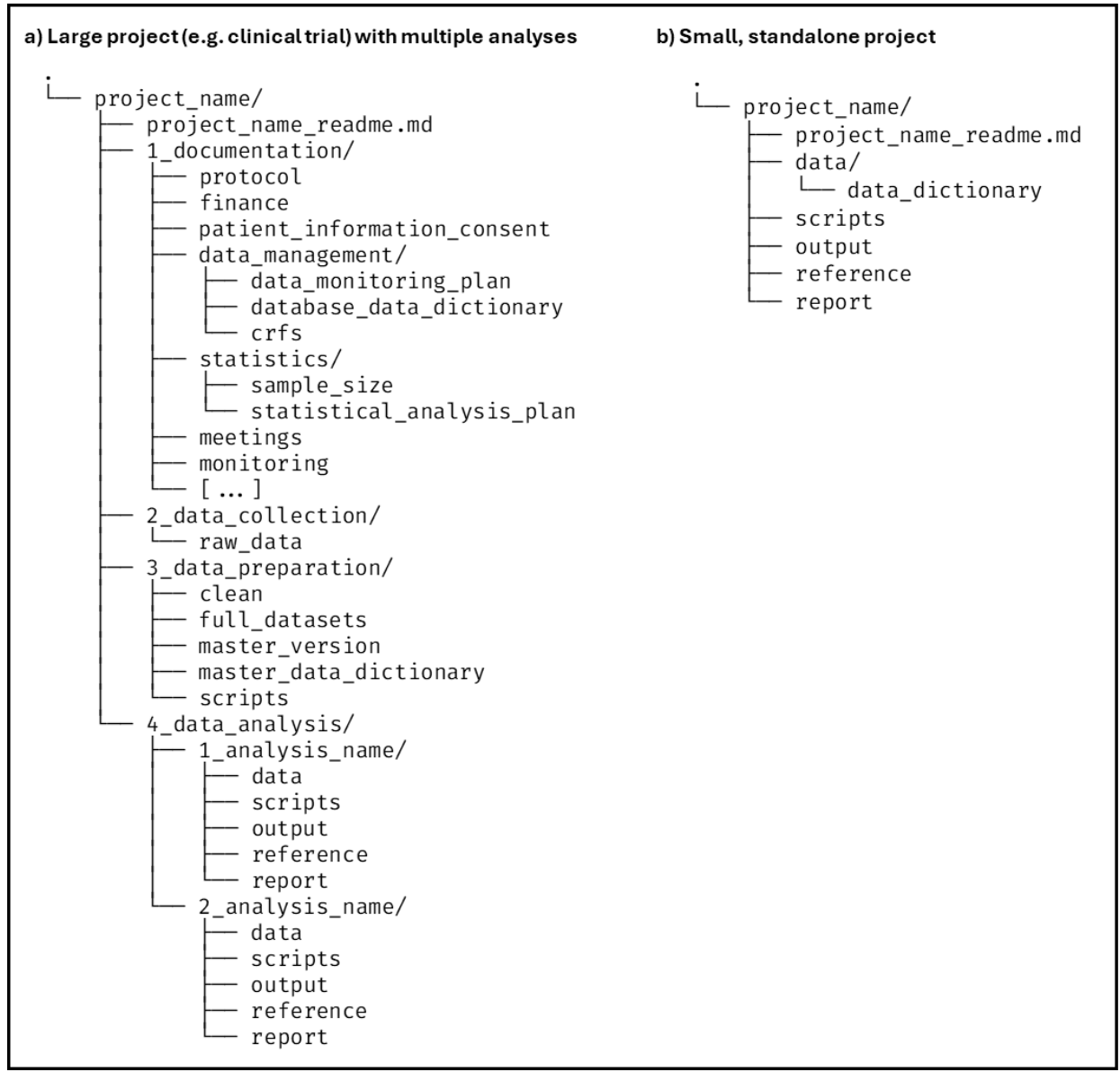